\documentclass[11pt,preprint]{aastex}

\slugcomment{in press ApJ Letters, 4 October 2006}

\begin{document}

\title{Detection of Methane on Kuiper Belt Object (50000) Quaoar}
\author{E.L. Schaller\altaffilmark{1}, 
M.E. Brown\altaffilmark{1}}
\altaffiltext{1}{Division of Geological and Planetary Sciences, California Institute
of Technology, Pasadena, CA 91125}
\email{schaller@caltech.edu}

\begin{abstract}

The near-infrared spectrum of (50000) Quaoar obtained at the Keck
Observatory shows distinct absorption features of crystalline water 
ice, solid methane and ethane, and possibly other higher 
order hydrocarbons.
Quaoar is only the fifth Kuiper belt object on which
volatile ices have been detected.
The small amount of methane on an otherwise water ice dominated
surface 
suggests that Quaoar is a transition
object between the dominant
volatile-poor small Kuiper belt objects (KBOs) and 
the few volatile-rich large KBOs such as Pluto and Eris.

\end{abstract}

\keywords{Kuiper belt --- planets and satellites}

\section{Introduction}

While once Pluto and Triton were the only objects in the outer solar system 
known to contain volatile ices on their surfaces, 
the recent discoveries of frozen methane on the 
large Kuiper belt objects (KBOs) Eris, Sedna, and 2005 FY9 have
shown that these objects are part of a new class of surface volatile rich
bodies in the outer solar system \citep{2005ApJ...635L..97B, 2006A&A...445L..35L,2005A&A...439L...1B, ebspec}.
In contrast to these bodies with detectable volatiles,
spectral observations of small KBOs over the past decade have 
found that most of these objects either contain varying amounts
of involatile water ice on their surfaces or have flat spectra with no 
identifiable features \citep{Kris}.
To understand the dichotomy between volatile rich and volatile free
surfaces in the outer solar system,
\citet{2007ApJ...659L..61S} constructed a simple model of 
atmospheric escape of volatile ices over the age of the solar system.
They found that while most KBOs 
are too small and hot to retain their initial volatile ices
to the present day, a small number are
large and cold enough to retain these ices on their surfaces.

As anticipated, the model suggests that the largest KBOs,
Eris, Pluto, and Sedna are all expected to 
retain surface volatiles, while the vast majority of the other known
objects in the Kuiper belt are expected to have lost all surface
volatiles. Two known intermediate-sized KBOs are predicted to be
in the transition region where they
may have differentially lost some volatile ices (N$_2$) but retained
others (CH$_4$). One of these transition objects,
2005 FY9, with a diameter of $\sim$1450 km \citep{2007astro.ph..2538S} 
does indeed appear to contain 
abundant CH$_4$ but be depleted in N$_2$.

The other object that appears to be in the
volatile non-volatile transition region is Quaoar, with a 
diameter of 1260$\pm 190$ km\citep{2004AJ....127.2413B}.
The infrared 
spectrum of Quaoar does not resemble that of 2005 FY9, however.
Quaoar's spectrum is dominated by absorptions due to involatile water ice,
which is not detected at all on 2005 FY9. In addition,
\citet{2004Natur.432..731J} reported the detection of an absorption
feature near 2.2 $\mu m$ 
that they attributed to ammonia hydrate.
They also detected the presence of crystalline water ice which, at the $\sim$ 40 K radiative equilibrium temperature of Quaoar, is
thought to be converted to amorphous water 
ice on a relatively short ($\sim 10$ Myr) 
timescale by cosmic ray bombardment.  
The crystallinity of the water ice and the
detection of the 2.2 $\mu m$ feature that they attributed to ammonia hydrate
led \citet{2004Natur.432..731J} to suggest that Quaoar may have experienced
relatively recent cryovolcanic activity.

In this paper, we present a new infrared spectrum of Quaoar with a 
signal-to-noise in the K-band six times greater than that of
\citet{2004Natur.432..731J}
and model the ices present on the surface.

\section{Observations}  

Near-infrared spectra of Quaoar were obtained on 12 July 2002,
and 23, 24 and 25 April 2007 using NIRSPEC,
the facility medium to high resolution 
spectrometer on the Keck telescope \citep{1998SPIE.3354..566M}.  
Three
separate grating settings were used to completely cover the region
from 1.4 to 2.4 $\mu m$.  
The 1.44-1.73 and 1.70-2.13 $\mu m$ regions were each covered in 
6 exposures of 200 seconds each 
and the 2.0 to 2.4 $\mu m$ region was covered in 82 exposures of 
300 seconds each.
Observations consisted of a series of exposures on two or three nod positions
along the 0.57'' slit.   
All observations were performed at an airmass of better than 1.5. 
Data reduction was carried out as described in \citet{2000AJ....119..977B}.
Telluric calibration was achieved by dividing the spectra by nearby G dwarf
stars observed at an airmass within 0.1 of that of the
target. 

Figure 1 shows the complete near-infrared spectrum of Quaoar.  
The absolute value
of the infrared albedo is obtained from the $R$ albedo of $0.092^{+0.036}_{-0.023}$
\citep{2004AJ....127.2413B}
, the V-R color of  $0.64\pm0.04$,
and an estimated V-J color of $2.1\pm0.2$ using typical values found by
\citet{2003Icar..161..501M}.
Errors in the overall absolute albedo
calibration are
of the order of 30\% and are dominated by the uncertainty in the optical
albedo.

The characteristic 
broad absorptions due to water ice are apparent at 1.5 and 2.0 $\mu m$.
In addition, the presence of the unique crystalline water ice feature at 1.65 $\mu m$ 
indicates that at least some of the water ice is in crystalline form.
We also detect the absorption feature at 2.2 $\mu m$ previously attributed to 
ammonia hydrate as well as a series of broad absorption
features beyond 2.25 $\mu m$.

\section{Spectral Modeling}
    
We first attempt to model the spectrum of Quaoar with a mixture
of water ice and a dark featureless material. 
Using the bidirectional reflectance models
of \citet{1993tres.book.....H} we model a spatially segregated mixture
of crystalline water ice grains \citep{1998JGR...10325809G} and a dark red material. 
Optical path length was parameterized as a grain size
in a scattering regolith.  In order to investigate the additional absorption
features longward of 2.1 $\mu m$ not due to water ice, we performed a least-squares
best fit to the 1.4 to 2.1 $\mu m$ region where the fractional 
abundance of the ice, the grain size of the ice, and the color of the dark
material were allowed to vary.
Figure 1a shows the best-fit model with
10 $\mu m$ water ice grain sizes, an ice fraction of 
40\% linearly mixed with a red continuum with a spectral slope of 10\% per $\mu m$
and an albedo of 13\% at 2 $\mu m$.
The spectrum could be equally well fit with a model where water ice was
significantly more abundant and the dark material was intimately mixed, 
rather than linearly mixed, with the ice.

The water ice model provides a good fit to the data from 1.4 to 2.1 $\mu$m, 
and, as expected, deviates in the region beyond 2.1 $\mu$m.
In order to explore the spectral shape of the deviation beyond 2.1 
$\mu m$, we show, in Figure 2a, the ratio of the Quaoar spectrum to the 
water ice model.
In addition to the 2.2 $\mu m$ feature seen by \citet{2004Natur.432..731J},
we see additional distinct but broader absorption 
features at 2.32 and 2.38 $\mu m$.  
The locations and depths of these three absorption
features as well as the general shape of the ratio spectrum beyond
2.1 $\mu$m are well fit by absorptions due to methane ice. 
A model methane spectrum using the laboratory data of 
\citet{2002Icar..155..486G} with 100 $\mu m$ grain sizes provides 
an excellent fit to the 2.0 - 2.4 $\mu$m ratio spectrum.
In addition, smaller absorptions at 1.8 and 1.72 $\mu m$ 
are also well fit by methane (Figure 1b).
We measured the
band center of the 2.2 $\mu m$ feature to be at 2.205$\pm$0.002 which is 
consistent with that expected from pure methane \citep{1997Icar..127..354Q}.
Hydrated ammonia, in contrast, has a single absorption in this region
at a wavelength between 2.21 and 2.24 $\mu m$, depending on the degree
of hydration \citep{2007Icar..190..260M}. While the Quaoar absorption feature at
2.205$\pm$0.002 is perhaps consistent with the 2.21 $\mu m$
wavelength of the absorption feature of the most hydrated ammonia, 
ammonia cannot explain any of the additional major absorptions seen on
Quaoar, all of which are well explained by the presence of methane. We
thus conclude that the absorption feature initially attributed to 
ammonia is actually one of a series of absorption features caused by
methane.

We can model the overall spectrum of Quaoar with a linear
mixture of 35\% water ice, 5\% methane, and 60\% dark material.
While most features of the spectrum are well fit by this model, 
additional
absorptions near 2.27 $\mu$m  still cannot be explained by methane (Figure 2b).
On 2005 FY9, \citet{ebspec} found that the surface contained not just 
volatile methane, but a small amount of involatile ethane, hypothesized
to derive from irradiation of the solid methane. The deviation from
the methane spectrum on Quaoar also appears to be well fit, at least in 
part, by the presence of ethane. Figure 2b shows a model of
10 $\mu m$ ethane grains constructed from the absorption coefficients
of \citet{1997Icar..127..354Q}. In particular, the two strongest
absorption features at 2.27 and 2.32 $\mu$m are clearly 
detected as they are on 2005 FY9.
An additional small absorption feature at 2.36 $\mu m$ 
remains unidentified but may be due to a higher-order hydrocarbon.

Figure 1b shows the complete model fit with a spatial mixture of
35\% water ice, 55\% dark material,
6\% methane and 4\% ethane. While 
the specific values of these parameters have little unique meaning, 
as many different values could give similar fits, in general we find
that methane and ethane are minority species on a water ice dominated
surface.  The slight increase in the albedo of our best-fit model in the 1.4-1.6 $\mu m$
region compared to the spectrum indicates that a single sloped continuum  
in the infrared is insufficient.
No evidence for other volatile species is detected; in 
particular, we
did not detect the 2.148 $\mu m$ N2 absorption feature
nor the 2.352 $\mu m$ CO absorption feature. 

\section{Discussion}

With significantly higher signal-to-noise in the 2.0 - 2.4 $\mu m$ region,
the 2.2 $\mu m$ absorption feature on Quaoar previously identified as
ammonia hydrate \citep{2004Natur.432..731J} is clearly seen to be due to
methane ice. No compelling evidence is seen for the presence of
ammonia. The presence of crystalline water ice on the surface of
Quaoar still
remains unexplained because it is expected that
ice should currently exist in the amorphous
form on the $\sim$ 40 K surface of Quaoar.
However, the presence of the 1.65 $\mu m$ absorption feature
due to crystalline water ice in the spectrum of
every well observed water ice rich KBO
(even down to diameters of only a few hundred kilometers) \citep{Kris}
suggests that exotic processes such as cryovolcanism are unlikely
to be required. The presence of crystalline
water ice on so many small outer solar system bodies 
may indicate that our current understanding of the physics of the
crystalline/amorphous phase transition may not be complete.
The spectrum of Quaoar is consistent with that of
a cold geologically dead object slowly losing the last of its volatile
ices by escape in a tenuous, perhaps patchy, atmosphere.

Ethane is an expected by-product of irradiation of methane ice
\citep{2003NIMPB.209..283B}. The presence of ethane on
Quaoar and on 2005 FY9 supports the suggestion of \citet{ebspec} that 
these irradiation products are preferentially seen on bodies with
large abundances of pure methane rather than on the bodies where the methane 
is diluted in nitrogen. Quaoar also appears to be rich in more complex
irradiation products.
Quaoar is the only water ice rich KBO which
has a red color in the visible. 
Other water ice rich KBOs like 
Orcus, Charon, and 2003 EL61 and its family of collisional fragments are all
blue in the visible \citep{Kris}.  
Quaoar's red surface is likely due to the continued irradiation of
methane, ethane, and their products on the surface \citep{2006ApJ...644..646B}.

While methane on Quaoar
 is sufficiently volatile that it is likely to seasonally 
migrate if Quaoar has a moderate obliquity, ethane and the other irradiation
products are essentially involatile at Quaoar's temperature. 
Quaoar is therefore likely to have an irregular
covering of irradiation products, perhaps leading to rotational variability
in its visible color and in the abundance of ethane. 
Continued observations of this object will provide
insight into the nature of the volatile non-volatile transition
and atmospheric escape in the outer solar system.

 {\it Acknowledgments:} 
We thank an anonymous referee for a helpful review.  
E.L.S. is supported by a NASA Graduate Student
Research Fellowship.  The data presented herein were obtained at the W.M.
Keck Observatory, which is operated as a scientific partnership among the California Institute of Technology, The University of California and the National 
Aeronautics and Space Administration.  The observatory was made
possible by the generous financial support of the W.M. Keck Foundation.

\clearpage

\begin{figure}
\plotone{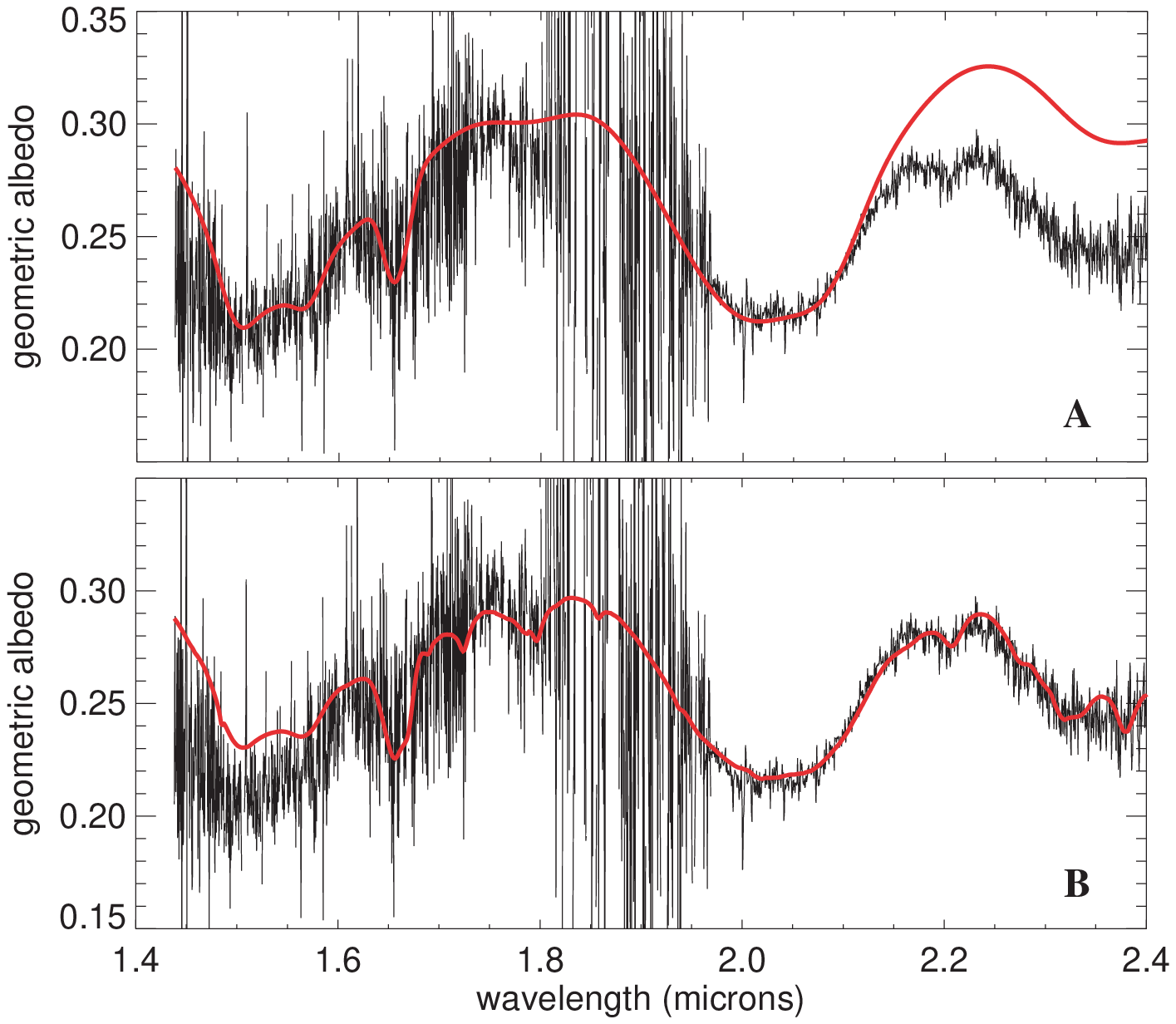}
\caption{A. Near-infrared spectrum of Quaoar obtained with NIRSPEC on the
Keck telescope \citep{1998SPIE.3354..566M}.  The signature of crystalline water ice is dominant. 
A best-fit water ice and dark red continuum model is shown.  The major
features of the spectrum are well fit by the model except for the 2.2-2.4
$\mu m$ region where additional absorption occurs.
B.  Quaoar spectrum shown with the best-fit water ice, continuum, methane 
and ethane model.  The presence of methane is required to fit the long 
wavelength end of the spectrum.}
\end{figure} 

\begin{figure}
\plotone{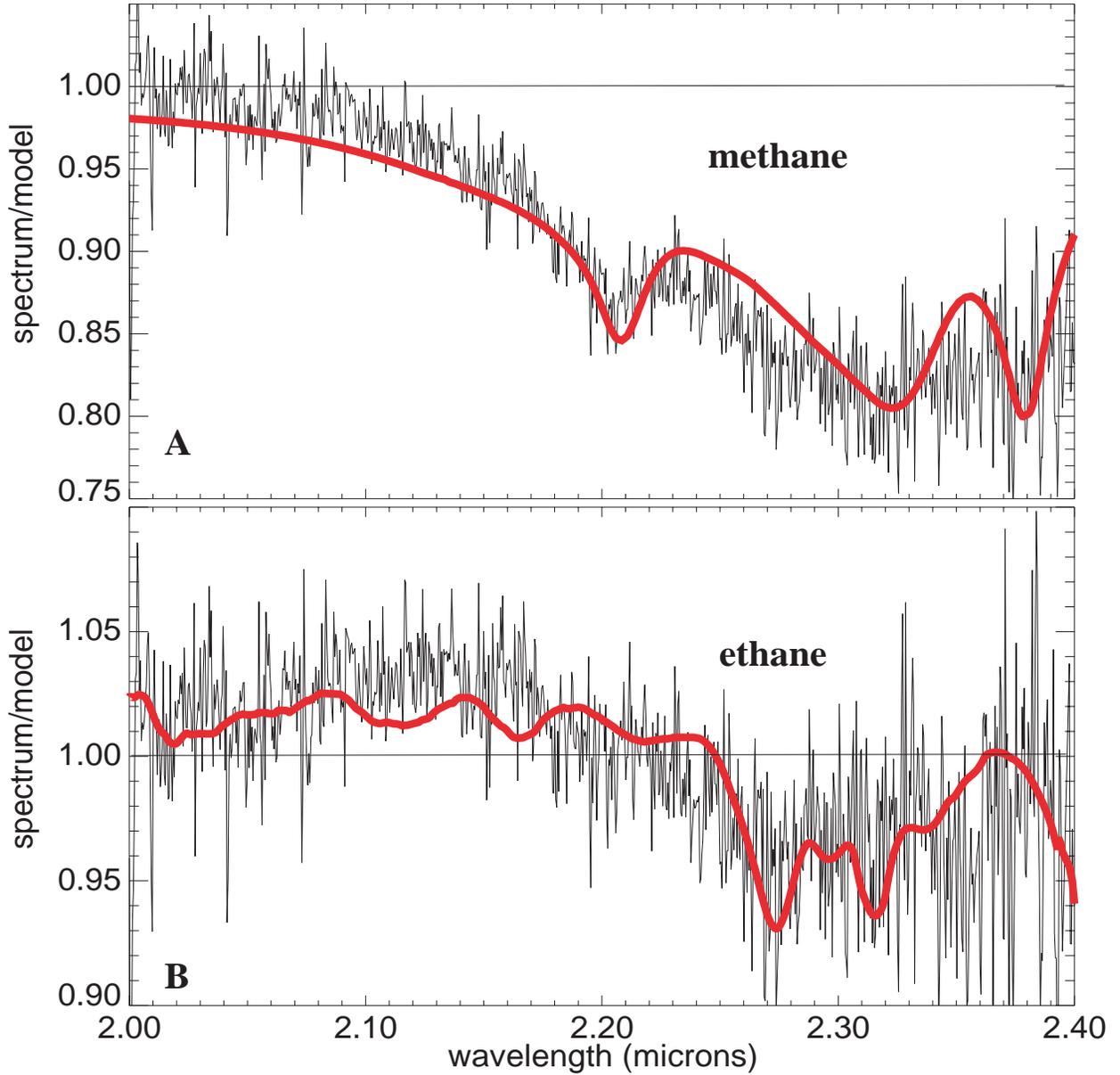}
\caption{A. Ratio of the spectrum to the best-fit water ice and red continuum model. In addition
to the 2.2 $\mu m$ feature seen by \citet{2004Natur.432..731J}, we see additional distict but 
broader absorption features at 2.32 and 2.38 $\mu m$ which are well matched
by methane.  A
model methane spectrum \citep{2002Icar..155..486G} with 100 $\mu m$ grain sizes provides 
an excellent fit to the ratio spectrum. 
B. Ratio of the spectrum to the best-fit water ice, continuum and methane model.
While most features of the spectrum are well fit by this model, additional 
absorptions occur near 2.3 $\mu m$.  The two strongest absorption features at 2.27 and 2.32
$\mu m$ are well fit by ethane.}
\end{figure}

\end{document}